\pdfoutput=1

\documentclass[fleqn,11pt]{article}

\usepackage{amsfonts,amsmath,amssymb}
\usepackage{latexsym}

\usepackage[dvipsnames]{color}

\usepackage{amsfonts,amssymb,cite}
\usepackage{graphicx}

\def\noi{\noindent}
\def\jnumber#1#2{\thispagestyle{empty} \noi\unitlength=1mm
    \begin{picture}(178,10)
            \put(177,15){\llap{\large\it Grav. Cosmol. No.\,#1, #2}}
                    \end{picture}}

\newcommand{\Title}[1]{\noi {{\Large\bf #1}}\\[1ex]}

\def\Aunames#1{\noi{\bf #1}}
\def\au#1{${}^{#1}$}
\def\Addresses#1{\medskip\noi \protect
\begin{description}\itemsep -3pt {\it #1} \end{description}}
\def\adr#1#2{\item[${}^{#1}$]{\it #2}}

\newcommand{\Abstract}[1]{\vskip 2mm \begin{center}
        \parbox{16.4cm}{\small\noi #1} \end{center}\medskip}

\def\email#1#2{\footnotetext[#1]{e-mail: #2}\addtocounter{footnote}{1}}

\def\nqq{\hspace*{-2em}}

\def\cm{\hspace*{1cm}}

\def\degC{\mbox{${}^\circ$C}}       
\usepackage{color}

\def\Acknow#1{\subsection*{Acknowledgments} #1}

\def\ConflictThey{\subsection*{Conflict of interest} 
The authors declare that they have no conflicts of interest.}

\def\Jl#1#2{#1 {\bf #2},\ }

\def\ApJ#1 {\Jl{Astroph. J.}{#1}}
\def\CQG#1 {\Jl{Class. Quantum Grav.}{#1}}
\def\DAN#1 {\Jl{Dokl. AN SSSR}{#1}}
\def\GC#1 {\Jl{Grav. Cosmol.}{#1}}
\def\GRG#1 {\Jl{Gen. Rel. Grav.}{#1}}
\def\IJMPD#1 {\Jl{Int. J. Mod. Phys. D}{#1}}
\def\JETF#1 {\Jl{Zh. Eksp. Teor. Fiz.}{#1}}
\def\JETP#1 {\Jl{Sov. Phys. JETP}{#1}}
\def\JHEP#1 {\Jl{JHEP}{#1}}
\def\JMP#1 {\Jl{J. Math. Phys.}{#1}}
\def\NPB#1 {\Jl{Nucl. Phys. B}{#1}}
\def\NP#1 {\Jl{Nucl. Phys.}{#1}}
\def\PLA#1 {\Jl{Phys. Lett. A}{#1}}
\def\PLB#1 {\Jl{Phys. Lett. B}{#1}}
\def\PRD#1 {\Jl{Phys. Rev. D}{#1}}
\def\PRL#1 {\Jl{Phys. Rev. Lett.}{#1}}

\def\lal{&&\nqq {}}

\def\beq{\begin{equation}}
\def\eeq{\end{equation}}
\def\bear{\begin{eqnarray}}
\def\bearr{\begin{eqnarray} \lal}
\def\ear{\end{eqnarray}}
\def\earn{\nonumber \end{eqnarray}}

\def\yy{\\[5pt] {}}

\begin{document}
\twocolumn[
\jnumber{issue}{year}

\Title{Estimating the Local Hubble Parameter from the Thermal Evolution \yy
of Earth and Mars
}

\Aunames{Yurii~V.~Dumin\au{a,b,1},
         Elizaveta~G.~Khramova\au{c},
         Ludmila~M.~Svirskaya\au{d,e}, and\\
         Eugen~S.~Savinykh\au{f}
}

\Addresses{
\adr a {Space Research Institute (IKI), Russian Academy of Sciences,\\
        Profsoyuznaya str.\ 84/32, Moscow, 117997 Russia}
\adr b {Sternberg Astronomical Institute (GAISh),
        Lomonosov Moscow State University,\\
        Universitetskii prosp.\ 13, Moscow, 119234 Russia}
\adr c {Pushkov Institute of Terrestrial Magnetism, Ionosphere and Radio Wave
        Propagation (IZMIRAN),\\
        Russian Academy of Sciences,
        Kaluzhskoe shosse 4, Troitsk, Moscow, 108840 Russia}
\adr d {South Ural State University (SUSU),
        Prosp.\ Lenina 76, Chelyabinsk, 454080 Russia}
\adr e {South Ural State Humanitarian Pedagogical University (SUSGPU),\\
        Prosp.\ Lenina 69, Chelyabinsk, 454080 Russia}
\adr f {Retired}
}


\Abstract
{The problem of local (\textit{e.g.}, interplanetary) Hubble expansion
is studied for a long time but remains a controversial subject till now;
and of particular interest is a plausible value of the local Hubble parameter
at the scale of the Solar system.
Here, we tried to estimate the corresponding quantity by the analysis of
surface temperatures on the Earth and Mars, which are formed by a competition
between a variable luminosity of the Sun and increasing radii of the planetary
orbits.
Our work employs
paleochemical and paleobiological data on the temperature of the ancient
Earth, on the one hand, and geological data on the existence of an ocean of
liquid water on the ancient Mars, on the other hand.
As follows from our analysis, the martian data impose only a weak
constraint on the admissible values of the Hubble parameter because of
the unknown salinity---and, therefore, the freezing point---of the martian
water.
On the other hand, the terrestrial data turn out to be much more valuable,
especially, for the Precambrian period, when temperature variation was
sufficiently smooth and monotonic.
For example,
in the framework of standard $\Lambda$CDM model with 70\% of dark energy,
contemporary value of the local Hubble parameter was found to be
70--90~km/s/Mpc under assumption that the Earth's surface temperature in
the end of Precambrian equaled~$ 45\degC $.
This is in reasonable agreement both with the intergalactic data and with
an independent estimate of the local Hubble parameter from tidal evolution
of the Earth--Moon system.}

\medskip

] 
\email 1 {dumin@pks.mpg.de, dumin@yahoo.com\\ \cm (Corresponding author)}

\section{Introduction}
\label{sec:Intro}

The question if the planetary orbits experience the universal Hubble
expansion is studied for over 90~years, since the well-known paper by
G.C.~McVittie~\cite{McVittie_33}, but remains a poorly understood subject
till now.
Two different points of view on this problem are summarized in
Table~\ref{Table1}.

The most of astronomers adhere to the criterion of gravitational binding,
namely, believe that Hubble expansion exists only in the
gravitationally unbound systems and disappears when the system becomes
gravitationally bound, \textit{i.e.}, its bodies experience a finite
(quasi-)periodic motion, such as stars in the galaxies or planets in
the planetary systems.
A well-known pictorial model of this situation is the inflating ball
(which represents the expanding space) with coins attached to its surface
(which mimic the galaxies)~\cite{Misner_73}.
So, from this point of view, the Hubble expansion should disappear when
a characteristic density of the system \textit{increases}.

\begin{table*}
\centering
\caption{\small
Comparison of two frequently-used criteria for the emergence/disappearance
of Hubble expansion.
\label{Table1}}
\bigskip
\begin{tabular}{lll}
\hline \\[-2ex]
\multicolumn{1}{c}{\bf Criterion of gravitational binding~\cite{Misner_73}}
&&
\multicolumn{1}{c}{\bf Einstein--Straus model~\cite{Einstein_45}}
  \\[0.5ex]
\hline \\[-2ex]
Hubble expansion exists only in the gravitationally-
&&
Rate of the local Hubble expansion is
  \\
unbound systems and disappears when motion of
&&
determined by Friedmann equation:
  \\
the gravitating bodies becomes confined.
&&
$  H = \sqrt{8 \pi G / 3} \;
       \sqrt{\rho_{\Lambda} + \langle \rho_{\rm m}\rangle} $.
  \\[3ex]
Therefore, the Hubble expansion \textit{decreases} when
&&
Therefore, the Hubble expansion \textit{increases}
  \\
a characteristic density of the system increases.
&&
with increase in the characteristic density.
  \\[0.5ex]
\hline
\end{tabular}
\end{table*}

On the other hand, there is also just an opposite point of view on this
problem.
Namely, according to the general-relativistic calculation by A.~Einstein
and E.G.~Straus~\cite{Einstein_45}, the Hubble expansion disappears when
the local background density of matter \textit{tends to zero}, and it is
restored again when the background density becomes equal to the standard
cosmological value.
This point of view turns out to be especially important in the context of
modern $\Lambda$CDM cosmological model: since the most fraction of the
background energy density is produced by the $\Lambda$-term (or `dark
energy'), which is distributed in space perfectly uniform, one can expect
that a substantial part of the Hubble expansion will survive even at
a very small (particularly, interplanetary) scale.

Unfortunately, although a few dozen of papers on the problem of local Hubble
expansion were published since the pioneering
papers~\cite{McVittie_33,Einstein_45}, almost all of them strongly depended
on the particular models employed by their authors; \textit{e.g.}
reviews~\cite{Bonnor_00,Dumin_16}. So, the ultimate answer can be given
only by the analysis of observational data.

One kind of observations suitable for this purpose might be
a tidal evolution of the Earth--Moon system~\cite{Dumin_02,Dumin_03}.
Namely, since a tidal bulge formed by the lunar attraction on the Earth's
surface is not directed exactly to the Moon but slightly shifted
due to non-zero relaxation time, there is a continuous exchange of angular
momentum between the orbital rotation of the Moon and proper rotation of
the Earth.
As a result, the Moon acquires the angular momentum and recedes from
the Earth with rate~$ \dot{R} $, while the Earth loses its proper angular
momentum and decelerates, resulting in the increase in the length of day
$ {\dot{T}}_{\rm E} $ (for more details, see for example Fig.~1
in~\cite{Dumin_03}).

The above-mentioned quantities are related to each other by formula:
\begin{equation}
\dot{R} = k \, {\dot{T}}_{\rm E} \, ,
\label{eq:Rdot_Tdot}
\end{equation}
where
$ k{=}1.81{\cdot}10^5 $\,cm/s.
Then, a secular increase in the length of day can be derived from
the available series of astrometric observations, which were accumulated
since the invention of telescopes (a compilation of such data for
approximately 350~years can be found, \textit{e.g.}, in
monograph~\cite{Sidorenkov_02}).
The resulting value turns out to be about
$ {\dot{T}}_{\rm E}\,{=}\,0.9{\cdot}10^{-5} $\,s/yr,
which leads to
$ \dot{R}\,{=}\,1.6 $\,cm/yr
(last column in Table~\ref{Table2}).

\begin{table*}
\centering
\caption{\small
Estimating the local Hubble parameter from evolution of the Earth--Moon system.
\label{Table2}}
\bigskip
\begin{tabular}{lllll}
\hline \\[-2ex]
&&
\multicolumn{1}{c}{\bf Direct measurement by}
&&
\multicolumn{1}{c}{\bf Indirect calculation from the}
  \\
&&
\multicolumn{1}{c}{\bf the lunar laser ranging}
&&
\multicolumn{1}{c}{\bf Earth's rotation deceleration}
  \\[0.5ex]
\hline \\[-2ex]
Effects taken
&&
(1)~tidal interaction
&&
(1)~tidal interaction
  \\
into account
&&
(2)~local Hubble expansion
&&
  \\[2.5ex]
Numerical value
&&
$ 3.8{\pm}0.1 $~cm/yr
&&
$ 1.6{\pm}0.2 $~cm/yr
  \\[0.5ex]
\hline
\end{tabular}
\end{table*}

On the other hand, the rate of increase of the lunar orbital radius~$ \dot{R} $
can be measured directly by the lunar laser ranging (LLR) due to deployment of
a few retroreflectors on the lunar surface in the course of Apollo and
Lunokhod missions in the early 1970's~\cite{Dickey_94}.
In the last four decades, the LLR is performed with accuracy at least 2--3\,cm
(and sometimes is claimed to be even better), which is quite sufficient
to measure the above-mentioned values of~$ \dot{R} $.

Surprisingly, the rate of increase of the lunar orbit measured by LLR turns out
to be~3.8\,cm/yr (middle column in Table~\ref{Table2}), which is considerably
greater than the value expected from the tidal interaction by
formula~(\ref{eq:Rdot_Tdot}).
In fact, this problem was recognized already in 1980's and 1990's;
and a number of additional effects were suggested to reconcile these two
values, such as a variable moment of inertia of the Earth, long-term
quasi-periodic variations in the length of day, \textit{etc}.
Unfortunately, non of these effects was able to resolve the above-mentioned
discrepancy convincingly.
So, it was suggested in our works~\cite{Dumin_02,Dumin_03} to attribute
the excessive rate~$ \Delta \dot{R} $ measured by LLR just to the local
Hubble expansion.
It is interesting that as early as 1970, \textit{i.e.}\ well before the precise
LLR measurements, L.A.~King already noticed that the numerical value of Hubble
expansion in the Earth--Moon system would be rather close to the tidal effects
and, therefore, could play a substantial role~\cite{King_70}.

Then, the contemporary value of discrepancy, 2.2\,cm/yr, should correspond to
the expansion with local Hubble parameter
\begin{equation}
H_0^{\rm (loc)} = 56 \pm 8~\mbox{(km/s)/Mpc},
\label{eq:H0loc_LLR}
\end{equation}
which is somewhat smaller than its intergalactic values, derived both from
the analysis of cosmic microwave background (CMB) as well as from
the Cepheid and SN\,Ia distance scales~\cite{Ryden_17}.

However, the above-written value can be reasonably corrected taking into
account the standard Friedmann equation:
\begin{equation}
H_0 = \sqrt{\frac{8 \pi G}{3}} \,
      \sqrt{\rho_{\Lambda} + \langle \rho_{\rm D0} \rangle} \, ,
\label{eq:H_Friedmann_eq}
\end{equation}
where
$ \rho_{\Lambda} $~is the effective density of the cosmological constant
($ \Lambda $-term), and
$ \langle \rho_{\rm D0} \rangle $~is the average density of non-relativistic
(dust-like) matter, both `dark' and visible.
So, one can assume that the value of Hubble parameter at large scales
is formed by both these terms; while at the local scales only the first of
them contributes to the cosmological expansion, because the ordinary matter
experiences the peculiar motions with respect to the cosmological background
due to Newtonian forces.

Then, it can be easily found that ratio of the global to local Hubble
parameters should be~\cite{Dumin_05,Dumin_08}:
\begin{equation}
\frac{H_0}{H_0^{\rm (loc)}} =
\bigg[ 1 + \frac{\Omega_{\rm D0}}{\Omega_{\Lambda}} \bigg]^{1/2} ,
\label{eq:H_ratio}
\end{equation}
where $ \Omega $ is, as usual, a relative density of the respective component
of the cosmological background.
Then, at the commonly-accepted values $ \Omega_{\Lambda}\,{=}\,0.7 $ and
$ \Omega_{\rm D0}\,{=}\,0.3 $, we get
$ {H_0}/{H_0^{\rm (loc)}}\,{\approx}\,1.2 $;
so that
\begin{equation}
H_0 = 67 \pm 10~\mbox{(km/s)/Mpc},
\label{eq:H0_recalc}
\end{equation}
which is in perfect agreement with modern estimates of the global Hubble
parameter.

To avoid misunderstanding, let us mention that the value of~$ H_0^{\rm (loc)} $
derived in our first paper~\cite{Dumin_03} was substantially less
than~(\ref{eq:H0loc_LLR}), namely, about 33\,km/s/Mpc; so that deviation
from the global Hubble parameter was substantial.
This was because we employed there a different rate of secular increase in
the length of day, $ {\dot{T}}_{\rm E}\,{=}\,1.4{\cdot}10^{-5} $\,s/yr.
This value was derived by some researchers from combination of telescopic
observations in the last 350~years with the historical data on ancient
and medieval eclipses~\cite{Stephenson_84}.
Unfortunately, the accuracy of ancient data is very doubtful, and their
inclusion into the analysis is hardly reasonable; for a detailed review of
this topic, see recent work~\cite{Maeder_21}.
So, starting from paper~\cite{Dumin_05} we preferred to employ the value
of~$ {\dot{T}}_{\rm E}\,{=}\,0.9{\cdot}10^{-5} $\,s/yr, derived solely from
observations in the telescopic era, and the agreement with the large-scale
cosmological data on~$ H_0 $ became much better.

It is the aim of the present work to estimate the local Hubble parameter
for two other systems, namely,  Sun--Earth and Sun--Mars.
Unfortunately, it is impossible to undertake exactly the same kind of analysis
as outlined above, because there are no high-accuracy ranging data, such as
LLR, for the Earth and Mars.
An alternative approach can be based on tracing a thermal evolution of
these planets over considerable time intervals (about 4~billion years, which
is comparable to their total lifetime).
If the local Hubble expansion really exists, then it would appreciably affect
the planetary thermal evolution and, thereby, it should be possible to
derive the corresponding Hubble parameter.

Why is it desirable to take into consideration the additional planets?
Firstly, this is because any attempts to derive the cosmological parameters
from the planetary evolution suffer substantially from a lot of planetological
uncertainties.
However, one can expect that these uncertainties will be averaged out if
the analysis is performed for a few planets with different physical properties,
and the resulting values will be sufficiently reliable.
Secondly, the analysis based on LLR measurements is constrained to
the present-day value of the Hubble parameter $ H_0 $, since the respective
data were collected for the period of only a few decades.
On the other hand, the analysis based on thermal evolution, which will be
presented in the next section, covers a period of over 4~billion years,
\textit{i.e.} about one third of the total age of the Universe.
Therefore, it should be possible to get information about variations of~$ H $
and other cosmological parameters at a considerable timescale.

\section{Thermal evolution of the Earth and Mars}
\label{sec:Therm_evol}

The principal conjecture that thermal evolution of the solar-system planets,
particularly the Earth, during their lifetime could be substantially affected
by the cosmological expansion of their orbits was put forward by
M.~K{\v{r}}{\'{\i}}{\v{z}}ek in 2012~\cite{Krizek_12}.
The primary aim of that work was to resolve the so-called faint young Sun
paradox; \textit{e.g.}, reviews~\cite{Feulner_12,Khramova_24} and references
therein.
Namely, according to the contemporary models of solar evolution, its luminosity
4~billion years ago was reduced approximately by 25--30\%~\cite{Gough_81}.
As a result, the Earth's surface temperature is expected to be much less
than today.
For example, the entire water would be frozen in the period earlier than
2--2.5~billion years ago, which would prevent emergence and development of
the biological life on the Earth as well as formation of the specific
geological structures attributed to the action of liquid water.

This problem was recognized already in the late 1950s~\cite{Schwarzschild_58},
and a number of various physical mechanisms were suggested by now to resolve
it, such as the greenhouse effect, geothermal sources of energy, a variable
albedo of the Earth's surface, \textit{etc.}
Unfortunately, none of them could give the ultimate solution.
So, yet another hypothesis suggested by
M.~K{\v{r}}{\'{\i}}{\v{z}}ek~\cite{Krizek_12} was that the reduced luminosity
of the Sun in the past could be compensated by a smaller orbital radius of
the Earth, which subsequently increased up to the present value due to
the local Hubble expansion; the same idea was subsequently developed by him
in collaboration with L.~Somer~\cite{Krizek_15}.
Starting from the assumption that the Earth's surface temperature should be
kept constant, K{\v{r}}{\'{\i}}{\v{z}}ek and Somer found that the required
local Hubble parameter should be about a half of its standard intergalactic
value.

However, a number of recent paleochemical and paleobiological findings (which
will be discussed in more detail below, in Sec.~\ref{sec:Observations})
suggest that temperatures of the early Earth might be much greater than its
contemporary value, up to~$ 70{-}80\degC $ (so that the faint young Sun
problem becomes even harder).
In the recent paper~\cite{Dumin_25} we tried to correct the
K{\v{r}}{\'{\i}}{\v{z}}ek--Somer hypothesis by taking into account
the increased temperature in the past.
As a result, a required value of the local Hubble parameter was estimated
to be roughly two times greater---rather than two times smaller---than at
the intergalactic scale.

It is the aim of the present work to extend this analysis in two directions:
Firstly, we shall take into account the additional observational data both
for the Earth and Mars.
Of course, the martian atmosphere is very different from the terrestrial one.
However, if the long-term trends in climates of both planets are governed
primarily by the solar variability and local Hubble expansion, then
the resulting behavior of their temperatures should be universal, and it is
important to check if this fact really takes place.
Secondly, our previous work~\cite{Dumin_25} was based on the model where
the  entire cosmological dynamics is governed solely by the dark energy
(the cosmological $ \Lambda $-term), so that the Hubble expansion was
purely exponential.
Of course, this is a rather crude assumption, and it will be interesting
to inspect more realistic situations.

\subsection{Theoretical model}
\label{sec:Theor_model}

The simplest model of a global planetary heat balance can be described by the
equation~\cite{Pollack_79}:
\begin{equation}
\pi R^2 (1{-}A) K = 4 \pi R^2 (1{-}{\alpha}) \, \sigma \, T_{\rm s}^4 \, ,
\label{eq:Therm_bal}
\end{equation}
where
$ R $~is the planetary radius,
$ T_{\rm s} $~is the temperature of its surface,
$ A $~is the planetary albedo (reflective capability) in the visible spectral
range (which involves the most part of the incident solar energy),
$ \alpha $~is the effective albedo of the planetary atmosphere in the infrared
range (in which the thermal radiation goes away),
$ K $~is the so-called solar constant, \textit{i.e.}, a flux of the solar
energy incident per unitary area of the surface, and
$ \sigma $~is the Stefan--Boltzmann constant.
The left-hand side of equation~(\ref{eq:Therm_bal}) describes a flux of
the solar energy absorbed by the planetary cross-section~$\pi R^2 $, while
the right-hand side represents the flux of thermal infrared radiation emitted
by the entire surface of the planet~$ 4 \pi R^2 $; and both terms are
corrected for the opacities in the corresponding spectral ranges.

The so-called `solar constant' is actually the function of time:
\begin{equation}
K(t) = \frac{\displaystyle L(t)}{\displaystyle 4 \pi r^2(t)} \, ,
\label{eq:Solar_const}
\end{equation}
since both the solar luminosity~$ L $ and distance from the Sun to
the planet~$ r $ should depend on time.
From here on, we shall assume that this orbital radius is merely proportional
to the cosmological scale factor~$ a $.
(In fact, a self-consistent treatment in the framework of General Relativity
shows that the relation between~$ r $ and~$ a $ might be much more
sophisticated~\cite{Dumin_20}; but we shall not go into these details in
the present article.)

Denoting the physical quantities at the present time $ t\,{=}\,0 $ by
subscript~`0' and assuming that both~$ A $ and~$ \alpha $ do not depend on
time, one can easily find from equations~(\ref{eq:Therm_bal})
and~(\ref{eq:Solar_const}) a temporal variation of the relative temperature:
\begin{equation}
T_{\rm s}(t) / T_{\rm s0} =
  \left[ L(t) / L_0 \right]^{1/4} \left[ a(t) / a_0 \right]^{-1/2} .
\label{eq:T_rel_gen_var}
\end{equation}

Function~$ L(t) $ is commonly derived from complex numerical solutions
of the equations of solar evolution, but the final results can be
interpolated by rather simple formulas, such as~\cite{Gough_81}:
\begin{equation}
L(t) = L_0 \bigg( \! 1 - \frac{2}{5} \, \frac{t}{t_{\rm S}}
  \bigg)^{\!\!\! -1} ,
\label{eq:L_t}
\end{equation}
where $ t_{\rm S} \approx 4.7{\cdot}10^9 $\,yr~is the age of the Sun.
(So, the luminosity becomes smaller at negative~$ t $.)

At last, a temporal behavior of the cosmological scale factor~$ a(t) $
is determined by the well-known Friedmann equation~\cite{Olive_16},
which can be conveniently written in the following form:
\begin{equation}
\bigg( \! \frac{{a}^{\prime}}{a} \! \bigg)^{\!\! 2} = \,
  {\Omega}_{\Lambda} + \,
  {\Omega}_{\rm D}\bigg( \! \frac{a_0}{a} \! \bigg)^{\!\! 3} .
\label{eq:Friedmann}
\end{equation}
Here, $ {\Omega}_{\Lambda} $ and $ {\Omega}_{\rm D} $~are the relative energy
densities of the $ \Lambda $-term (dark energy) and the non-relativistic
(dust-like) matter, both `dark' and visible; and the prime denotes
a differentiation with respect to the normalized time~$ \tau $:
\begin{equation}
\tau = H_0 \, t \, .
\label{eq:tau_def}
\end{equation}

Since we consider here only the late stage of the cosmological evolution,
contribution from the relativistic matter (radiation) to
equation~(\ref{eq:Friedmann}), which is proportional to~$ 1 / a^4 $, was
ignored.
Besides, following the commonly-accepted paradigm, we assume that
the Universe is spatially flat; so that the curvature term, proportional
to~$ 1 / a^2 $, was also discarded, and
$ {\Omega}_{\Lambda} + {\Omega}_{\rm D} = 1 $.

Therefore, the required function~$ a(t) $, appearing in
formula~(\ref{eq:T_rel_gen_var}), can be obtained by a numerical integration
of the equation:
\begin{equation}
{a}^{\prime} \equiv \frac{{\rm d} a}{{\rm d} \tau} = \,
  a \, \sqrt{(1 - {\Omega}_{\rm D0}) + \,
  {\Omega}_{\rm D0}\bigg( \! \frac{a_0}{a} \! \bigg)^{\!\! 3}} \, ,
\label{eq:a_tau}
\end{equation}
where
$ a(0) = a_0 $,
and we are interested only in the solutions at $ \tau < 0 $.
It is commonly accepted in the contemporary cosmology that
$ {\Omega}_{\rm D0} \approx 0.25{-}0.3 $.

Thereby, substituting the functions~$ L(t) $ and~$ a(t) $ given by
formulas~(\ref{eq:L_t}) and~(\ref{eq:a_tau}) to~(\ref{eq:T_rel_gen_var}),
we get a relative variation of the planetary temperature in the past.
It is important to emphasize that this variation is `universal'---the same
for any planet---provided that its contemporary value~$ T_{\rm s0} $ is
taken appropriately,
for example, 288\,K for the Earth and 218\,K for
Mars~\cite{Pollack_79}.
Finally, by drawing a set of theoretical curves $ T_{\rm s}(t) / T_{\rm s0} $
and confronting them with the available observational data on the planetary
temperatures in the past, one should be able to estimate a possible
presence and magnitude of the local Hubble expansion within the solar system.

\subsection{Observational data on the temperatures of the Earth and Mars}
\label{sec:Observations}

\begin{figure*}
\centering
\includegraphics[width=17.5cm]{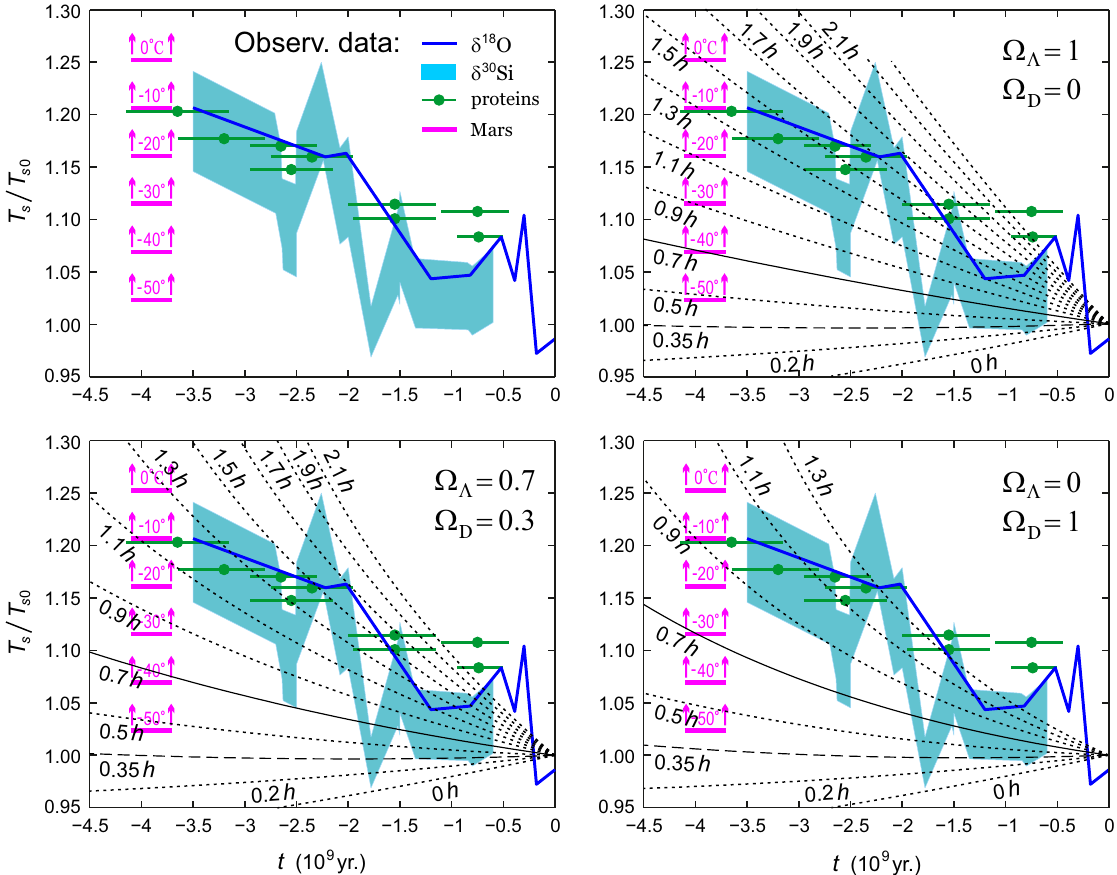}
\caption{\small
Observational data on the relative surface temperatures of the Earth and Mars
$ T_{\rm s}/T_{\rm s0} $ (upper left panel) \textit{vs.}\ the theoretical
predictions by a few cosmological models at various values of the Hubble
parameter normalized to $ h = 100 $~km/s/Mpc (dotted curves in three other
panels).
Solid curves correspond to the standard intergalactic value of the Hubble
parameter; and the long-dashed curves, to the original
K{\v{r}}{\'{\i}}{\v{z}}ek--Somer model~\cite{Krizek_12,Krizek_15}.}
\label{fig:T-t}
\end{figure*}

We shall employ below three major sets of experimental data on the planetary
paleotemperatures.
The first group of data is based on the concentrations of various
isotopes---first of all, $ {}^{18} \! $O and~$ {}^{30} \! $Si---deposited in
the minerals at the bottom of terrestrial oceans.
Since a relative variation in concentrations of different isotopes of
the same chemical element (\textit{e.g.}, ${}^{18} \! $O and~${}^{16} \! $O)
is sensitive to the temperature of the environmental water from which
the minerals were deposited, it can be used to extract the temperature in
the period of their formation.
Such studies started as early as 1960's and 1970's, and it was concluded
already in one of the first works~\cite{Knauth_76} that ``a few data for
the Precambrian suggest the possibility that Earth surface temperatures
may have reached about~$ 52\degC $ at 1.3~billion years and
about~$ 70\degC $ at 3~billion years [ago]''.
In other words, a climate of the early Earth might be significantly hotter
than today.

In our analysis, we shall use the newer data~\cite{Robert_06}, which are
potted in Fig.~\ref{fig:T-t} by the dark blue line
for~$ \delta\,{}^{18} \! $O
and by the light blue polygons for~$ \delta\,{}^{30} \! $Si.
As is seen, the scatter of data is substantial; however a general tendency
for the increasing temperature in the past is evident.

The second set of experimental data on the terrestrial paleotemperature
comes from the absolutely different branch of science---molecular biology
and genetics.
This is studying the temperatures of existence of the reconstructed protein
sequences in ancient bacteria~\cite{Gaucher_08}, which are shown in the same
figure by the dark green circles with error bars.
Again, one can see that, for the biological evolution to
proceed with a required rate, temperature on the early Earth
should be much greater than today.

At last, as regards the martian paleotemperatures, one can employ
a commonly-recognized fact---following from geological studies of this
planet, \textit{e.g.}~\cite{Li_25}---that a considerable portion of its
surface in the past was covered by the ocean and a network of
rivers~\cite{Gopalchetty_25}.
Therefore, the temperature in that period had to be above the freezing point
of water.
A temporal range of the corresponding period is estimated somewhat differently
by various authors, but the most frequently-cited time interval is 4.1 to
3.7~billion years ago.
(Let us mention that there are numerous reports on the dry river beds on
the martian surface that are dated by much later times, \textit{e.g.},
a few hundred million years ago or even younger; but they are usually
assumed to be fed by the underground thermal sources and, therefore,
not in equilibrium with the global climatic system.
So, they should not be interesting for our analysis.)

To impose constraints on the martian temperature from the existence
of liquid water, one should keep in mind two more circumstances:
Firstly, the atmospheric pressure on Mars in the past might be much
greater than its present-day value, which is about the triple point of
water, 611~Pa.
In principle, the variable pressure could shift the freezing temperature
appreciably for some substances.
Fortunately, this is not the case of water.
As can be seen in its phase diagram (\textit{e.g.}, Appendix in
paper~\cite{Zhang_15}), a boundary between the solid and liquid phases is
almost vertical (\textit{i.e.}, is fixed at the same temperature) within
a considerable pressure range, from the triple point (611~Pa) up
to~10~MPa.
Therefore, even if the atmospheric pressure on the ancient Mars would be
greater than its contemporary value by four orders of magnitude, this almost
did not affect the freezing point of water.

The second, more important item is the effect of salts dissolved in
the martian water on its freezing point.
If the water would be fresh (unsalted), then its freezing point should
equal~$ 0\degC $ or 273\,K.
However, it is well known that martian water can be strongly salted,
\textit{i.e.}, represent a brine~\cite{Gopalchetty_25}.
Therefore, its freezing point might be substantially reduced.
The freezing temperature of terrestrial oceans can be reduced down to
approximately~$ -2\degC $~\cite{Shuleikin_68};
but the ancient martian ocean might be much more salted and, therefore,
its freezing temperature reduced much stronger.
For example, the freezing point can be as low as~$ -21\degC $ for NaCl brine
and~$ -55\degC $ for CaCl${}_2$ brine at the optimal
concentrations~\cite{Chubik_70}.
Unfortunately, we do not have any information about a chemical composition
of the ancient martian ocean
and, therefore, can only consider a set of various options.
In Fig.~\ref{fig:T-t}, a series of different freezing temperatures---from
$ 0\degC $ to $ -50\degC $---is shown by the violet bars.
The upward arrows denote that a liquid ocean could exist above
the corresponding temperatures.
(Let us emphasize that temperatures of the Earth and Mars in this figure
are normalized to the different values:
$ T_{\rm Ms0} = 218\,K = -55\degC $ and $ T_{\rm Es0} = 288\,K = 15\degC $,
respectively~\cite{Pollack_79}.)

\subsection{Analysis for the entire lifetime of the Earth and Mars}
\label{sec:Entire_lifetime}

While the upper left panel in Fig.~\ref{fig:T-t} represents a summary of
observational data, three other panels demonstrate a confrontation of
these data with theoretical curves~(\ref{eq:T_rel_gen_var}) for three different
cosmological models, characterized by the relative densities of the `dark
energy'~$ {\Omega}_{\Lambda} $ and non-relativistic (dust-like)
matter~$ {\Omega}_{\rm D} $, both `dark' and visible.
(The Universe is always assumed to be spatially flat, so that
$ {\Omega}_{\Lambda} + {\Omega}_{\rm D} = 1 $.)
The theoretical dependences~$ T_{\rm s}(t) / T_{\rm s0} $ are drawn by
the black dotted curves, and the corresponding values of the contemporary
Hubble parameter~$ H_0 $ are written near each curve.
For concineses, these values are normalized to $ h = 100 $~km/s/Mpc.

In the case of commonly-accepted cosmological model with
$ {\Omega}_{\Lambda} = 0.7 $ and $ {\Omega}_{\rm D} = 0.3 $ (bottom left
panel), the best fit is achieved at
$ H_0 = 1.2{-}1.5~h = 120{-}150 $~km/s/Mpc, \textit{i.e.}, almost two times
greater than the standard value of Hubble parameter at the intergalactic
scales.
The martian data (violet bars) suggest that such values of~$ H_0 $ are
consistent with the assumption of weakly-salted ocean, whose freezing point
was above~$ -10\degC $.
Significant outliers in the terrestrial data at the late times (within
1~billion years from today) are not surprising: this period involved
an active biological evolution, resulting in the considerable variations of
coefficients~$ A $ and~$ \alpha $ in our model~(\ref{eq:Therm_bal}), which
were not taken into account.

Next, if we consider a hypothetical case of the cosmological model containing
solely the dark energy ($ {\Omega}_{\Lambda} = 1 $ and
$ {\Omega}_{\rm D} = 0 $, upper right panel), then the required values of
the Hubble parameter will be even greater:
$ H_0 = 1.4{-}1.7~h = 140{-}170 $~km/s/Mpc.

On the other hand, in the opposite case when the entire cosmological
background is formed by the non-relativistic (dust-like) matter
($ {\Omega}_{\Lambda} = 0 $ and $ {\Omega}_{\rm D} = 1 $, bottom right panel),
the best-fitting Hubble parameter can be somewhat reduced:
$ H_0 = 1.0{-}1.2~h = 100{-}120 $~km/s/Mpc.
Unfortunately, the corresponding theoretical curves become noticeably concaved
and fit the experimental data worse at the intermediate times.

The solid black curves in all panels refer to the standard intergalactic value
of the Hubble parameter, $ H_0 \approx 0.7~h $, and they are evidently
unsuitable for fitting the terrestrial data.
As regards the martian data, such value of~$ H_0 $ evidently requires the
strongly-salted ocean, whose freezing point is reduced down
to~$ -40\degC $.

At last, the long-dashed curves correspond to the
K{\v{r}}{\'{\i}}{\v{z}}ek--Somer case~\cite{Krizek_12,Krizek_15},
$ H_0 = 0.35~h $, when the increasing solar luminosity is well compensated by
the Hubble expansion and, thereby, the planetary temperature is kept constant.
This situation is evidently incompatible both with terrestrial and
martian data.

We shall not try here to recalculate the local Hubble parameter to the global
one, as we did in Sec.~\ref{sec:Intro} by introducing a 20\%~correction,
because the scatter of experimental points in Fig.~\ref{fig:T-t} turns out
to be comparable to or even greater than this quantity.

In summary, results of the present section are in general agreement with our
earlier findings~\cite{Dumin_25}, but here we performed the analysis
using an extended suite of observational data---including the martian
ones---and more carefully confronted them with various cosmological
models.

\subsection{Analysis for the Precambrian period of the Earth}
\label{sec:Precambrian}

\begin{figure}
\centering
\includegraphics[width=8.5cm]{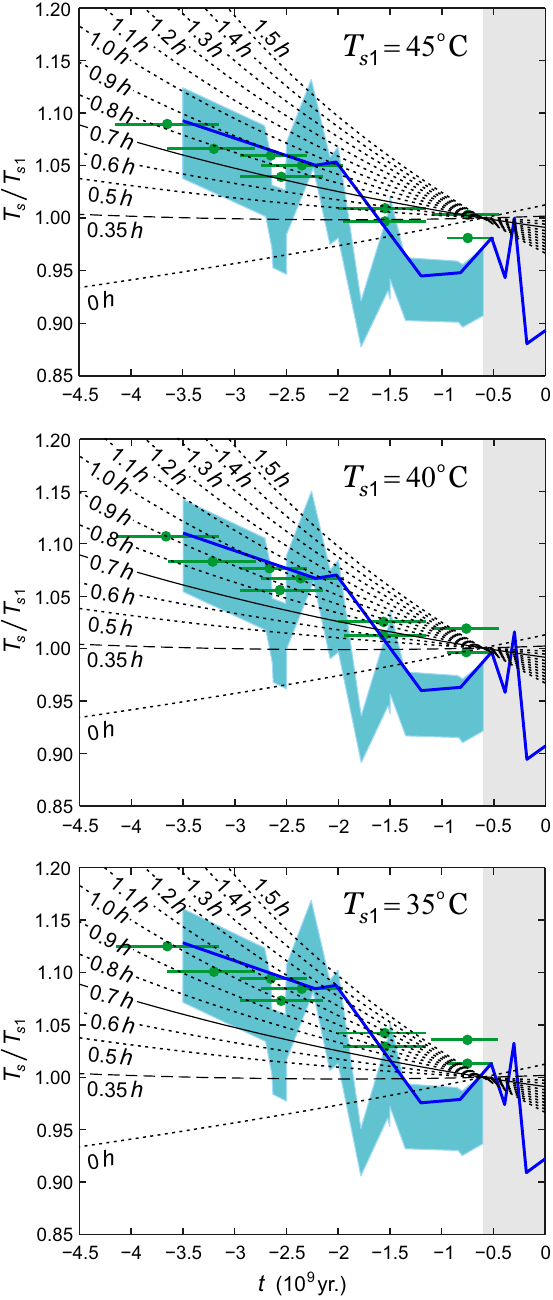}
\caption{\small
Observational data on the relative surface temperature of the
Earth~$ T_{\rm s}/T_{\rm s1} $ \textit{vs.}\ the theoretical predictions
by the standard cosmological model ($ {\Omega}_{\Lambda} = 0.7 $,
$ {\Omega}_{\rm D} = 0.3 $) for three different temperatures in the end of
Precambrian~$ T_{\rm s1} $.
All designations are the same as in Fig.~\ref{fig:T-t}, and the temporal
period excluded from the analysis is shaded in gray.}
\label{fig:T-t_Precambr}
\end{figure}

As was already mentioned in Sec.~\ref{sec:Entire_lifetime}, there are
significant outliers from the general trend at the sufficiently late
times (about 1~billion years from now).
In general, such outliers are not surprising: this was the period of active
biological evolution of the Earth, when both its atmospheric and surface
properties (characterized by the coefficients~$ \alpha $ and~$ A $,
respectively) could experience considerable variations.
Since temporal variations in~$ A $ and~$ \alpha $ were not taken into account
in our model~(\ref{eq:Therm_bal}), it might be reasonable to exclude
the corresponding period from the analysis and to normalize all temperatures
to the temperature in the end of Precambrian (approximately 600~million years
ago), which will be denoted by~$ T_{\rm s1} $.
Since martian data impose only a very weak constraint on the values of~$ H_0 $
(because of the unknown salinity of the martian water), they will not be used
here at all.

Following the same procedure as in Sec.~\ref{sec:Theor_model}, one can
derive the following expression for temporal variation of the relative
temperature:
\begin{equation}
T_{\rm s}(t) / T_{\rm s1} =
  \left[ L(t) / L(t_1) \right]^{1/4} \left[ a(t) / a(t_1) \right]^{-1/2} \! ,
\label{eq:T_rel_gen_var_Precamb}
\end{equation}
where
$ T_{\rm s1} \equiv T(t_1) $ and $ t_1 = -6{\cdot}10^8 $\,yr.
Functions $ L(t) $ and $ a(t) $ are calculated by the same methods as before.

Since temperature in the end of Precambrian~$ T_{\rm s1} $ is not
exactly known, this is an additional free parameter of the model.
According to the recent literature~\cite{Robert_06,Gaucher_08}, we can
estimate it to be in the interval from~$ 35\degC $ to~$ 45\degC $.
So, the results of our analysis are presented in Fig.~\ref{fig:T-t_Precambr}
for three different values of~$ T_{\rm s1} $.

As is seen in the upper panel, at $ T_{\rm s1} = 45\degC $
we get the \textit{contemporary} value of the Hubble parameter
$ H_0 = 0.7{-}0.9~h = 70{-}90 $~km/s/Mpc, \textit{i.e.}, in reasonable
agreement with intergalactic data (about 70~km/s/Mpc).
Unfortunately, the assumed temperature~$ 45\degC $ in the end of Precambrian
might be somewhat overestimated.

At the more realistic temperature $ T_{\rm s1} = 40\degC $, we get
$ H_0 = 0.85{-}1.0~h = 85{-}100 $~km/s/Mpc (middle panel), \textit{i.e.},
the agreement becomes a bit worse.

At last, at even lower temperature $ T_{\rm s1} = 35\degC $ (bottom
panel), the Hubble parameter is estimated to be
$ H_0 = 0.95{-}1.15~h = 95{-}115 $~km/s/Mpc, \textit{i.e.} the disagreement
increases further.
Anyway, when the analysis is performed only for the Precambrian period,
a general coincidence with the intergalactic data turns out to be much better
than for the entire lifetime of the Earth (Sec.~\ref{sec:Entire_lifetime}).

\section{Discussion and conclusions}

As follows from our consideration,
the most accurate approach to testing a presence of the local Hubble
expansion within the solar system is the analysis of tidal evolution
of the Earth--Moon system, which was briefly outlined in the Introduction.
This method gives the local Hubble parameter equal to~$ 56{\pm}8 $~km/s/Mpc,
which can be well recalculated to the standard intergalactic values,
about~70~km/s/Mpc, under assumption that it is formed locally only by
the perfectly-uniform `dark energy', while the irregularly-distributed forms
of matter produce an additional contribution at the larger scales.

The approach based on the thermal evolution of planets, at the first sight,
has an important advantage, because it enables one to trace the cosmological
dynamics over a considerable time interval (about 4~billion years).
Unfortunately, this method suffers from the much larger geophysical and
planetological uncertainties (\textit{e.g.}, due to the so-called `greenhouse
effect', a variable albedo of the planetary surfaces, \textit{etc.}).
Just these uncertainties are the most probable reason
of noticeable deviations of the local Hubble parameter from the global
one.

Besides, it is somewhat strange why the local values of Hubble parameter
derived in Sec.~\ref{sec:Therm_evol} were systematically larger than at
the intergalactic scale,
while just the opposite situation took place in the Earth--Moon system.
However, such
enhanced values do not represent any significant contradiction to
the theoretical concepts: the 20\% reduction given by
formula~(\ref{eq:H_ratio}) is actually the lower limit
on the local Hubble parameter.
In principle, the Hubble parameter
$ H = \sqrt{8 \pi G / 3} \:
      \sqrt{\rho_{\Lambda} + \langle \rho_{\rm m} \rangle} $
can be locally even greater than at the global scale due to a larger
average contribution from the non-relativistic (dust-like)
matter~$ \langle \rho_{\rm m} \rangle $.
Unfortunately, the problem of averaging the cosmological background density
remains a poorly understood subject till now, and we shall refrain here from
the particular estimates.

In fact, the most of geophysicists and planetologists believe that
thermal anomalies in the history of Earth and Mars should be explained by
the `local' physical processes, without any `external' cosmological effects.
However, let us emphasize once again that the concept of local Hubble
expansion, discussed in the present work, does not require any `new physics'
and is based solely on a more careful treatment of the commonly-accepted
equations of General Relativity and the standard cosmological models.
Moreover, a reasonable agreement between the values of Hubble parameter
obtained by a few absolutely different methods, from our point of view,
suggests that the undertaken analysis is meaningful.

In summary, we believe that further investigation of the problem of local
Hubble expansion should be important topic both for cosmology and
planetary sciences.

\Acknow{
YVD is grateful to
Yu.V.~Baryshev,
M.L.~Fil'chen\-kov,
V.~Gueorguiev,
I.D.~Karachentsev,
S.M.~Kopei\-kin,
M.~K{\v{r}}{\'{\i}}{\v{z}}ek,
A.~Maeder,
M.~Nowakowski, and
A.V.~Toporensky
for numerous discussions of the problem of local Hubble expansion.
We are also grateful to
O.~Primina and
A.~Truskov
for the support and encouragement of this work.}

\ConflictThey



\end{document}